\documentstyle[sprocl, amssymb]{article}

\def\Journal#1#2#3#4{{#1} {\bf #2}, #3 (#4)}

\def\NPB{{\em Nucl. Phys.} B}
\def\PLB{{\em Phys. Lett.}  B}
\def\PRL{\em Phys. Rev. Lett.}
\def\PRD{{\em Phys. Rev.} D}

\def\be{\begin{equation}}
\def\ee{\end{equation}}
\def\bea{\begin{eqnarray}}
\def\eea{\end{eqnarray}}

\newcommand {\nc}{\newcommand}
\nc{\eq}{\begin{equation}}
\nc{\en}{\end{equation}}
\nc{\norm}[1]{||{#1}||}
\def\intg{{\mathbb Z}}

\def\Drc{{\mathcal D}}

\def\ot{\otimes}

\nc{\slsh}[2]{{#1\mkern-#2mu/}}

\begin{document}

\title{SPONTANEOUS SYMMETRY BROKEN CONDITION IN (DE)CONSTRUCTING DIMENSIONS FROM NONCOMMUTATIVE GEOMETRY}

\author{JIAN DAI, XING-CHANG SONG}

\address{Institute of Theoretical Physics, School of Physics, Peking University\\
   Beijing, P. R. China, 100871\\E-mail: jdai@mail.phy.pku.edu.cn,
   songxc@ibm320h.phy.pku.edu.cn}




 \maketitle
 \abstracts{
 In this short report, a brief
 introduction to Arkani-Hamed, Cohen, Georgi model
 (ACG-model, {\it (de)constructing dimensions model}),
 whose main characters are that extra-dimensional space-time are generated
 dynamically from a four-dimensional gauge
 theory and that extra dimensions are lattices, will be given
 first. Then after a concise review of NCG on cyclic
 groups, actions for gauge fields along extra dimensions will be constructed by virtue of NCG and classical (vacuum)
 solutions will be solved, with low energy phenomenology being classified accordingly.
 As a conclusion, the behavior of
 spontaneous symmetry broken within ACG-model can be determined by noncommutative Yang-Mills theory.
}

\section{Arkani-Hamed, Cohen, Georgi (ACG) model: (de)constructing dimensions}
 \subsection{Motivations}
  The intuition of the existence of some silent extra dimensions
  has been renewed time to time; see literatures listed in \cite{Extra} for example.
  However as a quantum field theory in its usual formalism,
  this idea suffers the problem of dimensional couplings; namely,
  the perturbative series is \underline{out of control} once the energy scale
  approaches upwards the cutoff. One outlet, being reasonable from
  both technical and physical senses, is to {\it complete} the
  higher-dimensional theory with a more fundamental theory at its ultraviolet end, with or without gravity being
  included.
  To be another candidate of these {\it UV-completions},
  ACG-model \cite{ACG} is characterized as being fully \underline{under
  control}; in out understanding, this feature is the initial, if
  not unique, motivation of this speculation from ACG.
 \subsection{Pictures}
  Suppose that the {\it Platonic} physics without including gravity is a
  four-dimensional quantum gauge theory. The gauge group is of a
  direct product form
  $(\Pi_{i=0}^{N-1} G^i)\ot (\Pi_{i=0}^{N-1} G^i_s)$ where $N$ is a generic (large) integer.
  A collection of Weyl fermions transform according to
  $\chi^i\sim R_{def}(G^i)\ot\overline{R_{def}(G^i_s)}, \xi^i\sim
  R_{def}(G^i_s)\ot\overline{R_{def}(G^{i+1})}$, $i=0,1,2,...,N-1$
  in which $R_{def}(G)$ is the definition
  representation of $G$ and $\overline{R}$ is the complex conjugate representation of a representation $R$.
  Note that $i+1$ is computed modulo $N$, which is taken to be a convention henceforth in corresponding context.
  Let $G^i=U(m)$, $G^i_s=SU(n)$ for all $i=0,1,2,...,N-1$
  with generic $m, n$, where the choice of $G^i$ here is enlarged compared with the original ACG's consideration.
  A cyclic symmetry is assumed to set gauge couplings of $G^i_s$ to be $g_s$ and those of $G^i$
  to be $g_d$ and $g$ for the center subgroups and the quotients respectively.
  $g_d$ is supposed to be weak for all energy scales that we are interested.
  By the mechanism of {\it dimensional transmutation}, the gauge couplings $g$, $g_s$ can be described as well by two energy
  scales $\Lambda$, $\Lambda_s$ respectively; here it is required that
  $\Lambda_s\gg\Lambda$.

  When energy scale is lower downwards $\Lambda_s$, gauge interactions of $G^i_s$ become strong and fermion
  condensates happen in a way that $\langle\chi^i\xi^i\rangle\sim 4\pi f_s^3
  U_{i,i+1}$
  where $f_s$ is a {\it decay constant} of mass dimension and
  $U_{i,i+1}$ is $\sigma$-model scalar transformed as
  $U_{i,i+1}\sim R_{def}(G^i)\ot\overline{R_{def}(G^{i+1})}$.
  Below, let $m=1$ for simplicity hence the subscripts ``d'' are
  omitted. The low energy effective theory is read as
  \eq\label{Eff}
   S=\int d^4x \sum_{i=0}^{N-1} (-{1\over{4g^2}}F^i\cdot F^i+f_s^2 (D^\mu U_{i,i+1})^\dag D_\mu
   U_{i,i+1})+(\mbox{irrelevant terms})
  \en
  where the covariant derivative is defined to be $D_\mu U_{i,i+1}=\partial_\mu U_{i,i+1}-iA^i_\mu U_{i,i+1}
  +iU_{i,i+1}A^{i+1}_\mu$. {\bf The center observation of ACG is that action Eq.(\ref{Eff}) is
  on the other hand a five-dimensional gauge theory where the
  fifth direction is a periodic lattice with $N$ sites}. The following geometric and physical quantities
  can be computed, as low energy phenomenology provided by
  ACG-model.
  Lattice spacing $a=1/(\sqrt{2}gf_s)$, circumference $R=Na$ and five
  dimensional gauge coupling $g_5=ag^2$.
  The fluctuations $U_{i,i+1}$ {\it higgs} gauge
  symmetry down to a diagonal subgroup and mass square spectral of gauge
  bosons are
  \eq\label{MK}
   M_k=8g^2f_s^2sin^2(\pi k/N)
  \en
  where $k=0,1,\ldots ,N-1$, and
  diagonal gauge coupling is $g_4^2=g_5^2/R$.

  The above-mentioned observation also triggered our intuition to
  utilize noncommutative geometry (NCG) method to study scalar potential
  in ACG-model because the dynamics of Yang-Mills field is {\it
  geometric} and the (differential) geometry of lattices is a
  specification of NCG. In fact, the first attempt to adopt
  NCG in ACG-model was explored by M. Alishahiha
  \cite{Alishahiha} whose concern is concentrated mainly on gravity.
\section{Noncommutative Yang-Mills theory on cyclic groups}
 \subsection{Noncommutative geometry on cyclic groups}
  A comprehensive introduction to NCG can be found in \cite{NNCG}.
  NCG on a cyclic group can be characterized by a generalized {\it Dirac
  operator}. Let $\intg_N$ be $N$-order cyclic group, then Dirac
  operator on $\intg_N$ is defined to be
  $\Drc[\omega]=\bar{\omega}Tb^\dag+T^\dag b\omega$
  in which $T$ is the induced translation acting on functions over $\intg_N$, $b,b^\dag$ is a pair of fermionic
  annihilation/creation operators, and $\omega$ is
  so-called {\it link variable} being parametrized by $\omega=\rho
  e^{i\theta}$. Under a gauge transformation $U$, which is a unitary
  function on $\intg_N$, $\omega\rightarrow (TU)\omega U^\dag$.
  Two gauge covariants, which are the candidates of generalized field strength, are defined by
  \eq
   F_1:=\Drc[\omega]^2
   =T^\dag(\phi)bb^\dag+\phi b^\dag b,
   F_2:=\partial^\dag(\phi)b\wedge b^\dag=:\Drc[\omega]\wedge\Drc[\omega]
  \en
  in which $\phi:=\rho^2$, and the formal partial derivative $\partial:=T-Id$.
 \subsection{Dynamics of Dirac operator}
  Thanks for ACG's observation, the following identifications can be carried out.
  Firstly, the $N$-site lattice generated dynamically in ACG-model is identified with
  $\intg_N$; secondly, $U_{i,i+1}$ is identified with $\omega$; last and most importantly, scalar potential
  in action ~(\ref{Eff}) is identified with Yang-Mills action on
  $\intg_N$. Below three types of Yang-Mills actions will be
  constructed.
  \subsubsection{$V_I=Tr(F_1^2)$}
   After a representation for $b, b^\dag$ being specified,
   $V_I\sim\sum_{i\in\intg_N}\phi(i)^2$, whose equation of motion admits only zero solution obviously;
   consequently, no spontaneous symmetry broken happens under
   this construction.
  \subsubsection{$V_{II}=Tr(F_2^2)$}
   Introduce a lattice {\it laplacian}
   $\bigtriangleup=\partial\partial^\dag$, then
   \[
    V_{II}\sim \sum_{i\in\intg_N}\phi(i) (\bigtriangleup\phi)(i)
   \]
   Equation of motion is read as
   \[
    \rho\bigtriangleup\phi=0
   \]
   whose nontrivial solutions are constant functions.
   This fact can be understood schematically by the {\it extremal value principle} in commutative harmonic
   analysis \cite{MPEQ}. This class of solutions describes {\it
   flat directions}, because for any of them the potential energy
   equals to zero.
  \subsubsection{$V_{III}={1\over 2}Tr(F_2^2 +{\alpha\over 4}F_1^2-{\beta\over 2}F_1)$}
   where the coupling constants $\alpha\geq 0$. Note that the last
   term in $V_{III}$ is called {\it Sitarz term} \cite{Sitarz},
   which vanishes in continuum limit if there exists such a limit.
   The equation of motion is given by
   \[
    \rho(2t\phi-T(\phi)-T^\dag(\phi)-\beta)=0
   \]
   where $t:=(2+\alpha)/2\geq 1$. In any circumstance, there exists at least a zero solution.
   It is easy to see that if $\beta<0, \alpha\geq 0$ or $\beta=0,\alpha>0$, the zero solution is
   the minimum of $V_{III}$, which implies that no SSB happens.
   So below only case $\beta>0, \alpha\geq 0$ will be considered.

   \underline{Translation invariant solution}
   (ACG phase) $\phi(i)={\beta\over \alpha}, \forall
   i\in\intg_N, \alpha >0$; no solution exists for $\alpha=0$ in
   this case.
   Energy is computed to be $V_{III}=-N\beta^2/\alpha=:V^{inv}_N$.
   Accordingly, the same ACG type of SSB pattern is found
   out as shown in Eq.(\ref{MK}).

   \underline{Translation non-invariant solution}
   (nontrivial phase) $\phi=\beta v$,
   \[
    v(i)=\sum_{j=0}^{N-2-i}f_{i,i+j}(t)\psi_{i+j}(t),
    i=0,1,...,N-2; v(N-1)=0
   \]
   in which $f_{i,j}(t)=\frac{U_i(t)}{U_j(t)}$,
   $\psi_n=\frac{\Sigma_n(t)}{U_{n+1}(t)}$,
   $\Sigma_n(t)=\sum_{j=0}^nU_j(t)$. $U_n(t)$ are {\it Chebyshev's polynomials
   of the second kind} whose roots are all real and lie in
   $(-1,1)$; for all $n$, $U_n(t)>0$ if $t\geq 1$ \cite{Table}.

   {\it Energy of low energy phenomenology of nontrivial phase}, $\alpha>0$. For $N=3$,
   $V_{III}=-\frac{\beta^2}{2+2\alpha}>V^{inv}_3$. For $N=4$,
   \[
    V_{III}=-\frac{{3\over 4}\alpha^3+{11\over
    2}\alpha^2+{23\over 2}\alpha+5}{\alpha^2+4\alpha+2}\beta^2
    \left\{
    \begin{array}{ll}>V^{inv}_4,&\alpha\in(0,(\sqrt{73}-5)/3)\\
    =V^{inv}_4,&\alpha=(\sqrt{73}-5)/3\\<V^{inv}_4,&\alpha>\sqrt{73}-5)/3
    \end{array}
    \right.
   \]
   {\it Gauge boson mass square matrix} is given by
   $M^2\sim$\[
    \left(
     \begin{array}{ccccccc}
      \phi(0)&-\phi(0)&0&0&\ldots&0&0\\
      -\phi(0)&\phi(0)+\phi(1)&-\phi(1)&0&\ldots&0&0\\
      0&-\phi(1)&\phi(1)+\phi(2)&-\phi(2)&\ldots&0&0\\
      \ldots &\ldots &\ldots &\ldots &\ldots &\ldots&\ldots\\
      0&0&0&0&\ldots&\phi(N-3)+\phi(N-2)&-\phi(N-2)\\
      0&0&0&0&\ldots&-\phi(N-2)&\phi(N-2)
     \end{array}
    \right)
   \]
   which is semi-positive-define and has only one zero mode. For
   $N=3$, the eigenvalues are $m^2\sim (0, 1,
   3)\frac{\beta}{1+\alpha}$ vs $(0,3,3)\beta/\alpha$ for ACG phase according to Eq.~(\ref{MK}); for $N=4$,
   $m^2\sim (0,2,1+r+\sqrt{1+r^2},1+r-\sqrt{1+r^2})\phi(0)$
   where $\phi(0)=\beta(\alpha +3)/(\alpha^4 +4\alpha^2 +2)$,
   $r=1+1/(3+\alpha)$, vs $(0, 2,2,4)\beta/\alpha$ for ACG phase.
\section{Conclusion}
   In general, different actions from NCG give rise to different low energy phenomena.
   With $V_I$, no SSB happens; with $V_{II}$, there appears a random balance;
   for ACG phase from $V_{III}$, scenario of ACG is recovered; nontrivial phase from
   $V_{III}$ is
   a distinctively novel pattern, which will be the real vacuum in
   strong coupling range of $\alpha$ when $N=4$.
\section*{Acknowledgments}
    This work was supported by Climb-Up (Pan Deng) Project of
    Department of Science and Technology in China, Chinese
    National Science Foundation and Doctoral Programme Foundation
    of Institution of Higher Education in China.

\end{document}